\documentclass[10pt,twocolumn,twoside]{IEEEtran}
\usepackage{latexsym}
\usepackage{amsmath}
\usepackage{amssymb}
\usepackage{algorithmic}
 \usepackage[ruled,linesnumbered,noend,algo2e]{algorithm2e}
\usepackage{algorithm}
\usepackage{epsfig}
\usepackage{cite}
\usepackage{algorithmic}
\usepackage{overpic}
\usepackage{multirow}
\usepackage[table]{xcolor}
\usepackage{graphicx}
\usepackage{colortbl,array}
\usepackage{multirow,bigdelim}
\usepackage{graphicx}
\usepackage{subfigure}
\usepackage{booktabs}
\usepackage{booktabs,amsmath}
\usepackage{fancyhdr, graphicx, amsmath, amssymb}
\usepackage{nomencl}

\newtheorem{theorem}{Theorem}

\newtheorem{assumption}[theorem]{Assumption}

\usepackage{url}
\ifCLASSINFOpdf
\else
\fi
\newcommand{\RNum}[1]{\uppercase\expandafter{\romannumeral #1\relax}}

\begin{document}

\title{On the Impact of Bounded Rationality in Strategic Data Gathering} 

\author{Anju~Anand and Emrah~Akyol
\thanks{A.~Anand and E.~Akyol are with the Department of Electrical and Computer Engineering, Binghamton University--SUNY at Binghamton, NY, 13902 USA (e-mail: \{aanand6, eakyol\}@binghamton.edu).}
\thanks{This research is supported by the NSF via grants CCF \#1910715 and CAREER \#2048042.}
\thanks{This paper will be presented in the 5th IFAC Workshop on Cyber-Physical Human Systems.}
}

\maketitle

\begin{abstract}
We consider the problem of estimation from survey data gathered from strategic and boundedly-rational agents with heterogeneous objectives and available information.  Particularly, we consider a setting where there are three different types of survey responders with varying levels of available information, strategicness, and cognitive hierarchy: i) a non-strategic agent with an honest response, 
ii) a strategic agent that believes everyone else is a non-strategic agent and that the decoder also believes the same, hence assumes a naive estimator, i.e., level-1 in cognitive hierarchy
iii) and strategic agent that believes the population is Poisson distributed over the previous types, and that the decoder believes the same. 
We model each of these scenarios as a strategic classification of a 2-dimensional source (possibly correlated source and bias components) with quadratic distortion measures and provide a design algorithm. Finally, we provide our numerical results and the code to obtain them for research purposes at  \url{https://github.com/strategic-quantization/bounded-rationality}.
\end{abstract}
\IEEEpeerreviewmaketitle

\section{Introduction}
Consider the following scenario involving a survey designed to gauge public reception of a new plastic product, with responses influenced by respondents' attitudes toward climate change. Respondents' scores range from 1 (`will definitely not use') to 4 (`will definitely use'), and the survey needs to account for potential biases as well as varying levels of rationality among respondents. We model this problem using the hierarchical cognitive type model as studied by \cite{camerer2004cognitive}, considering three types of respondents:
\begin{itemize}
\item Type 0 (Honest-Nonstrategic Respondents): These respondents provide truthful information based on their actual opinions about the product, unaffected by their considerations of climate change or any desire to bias the survey.
\item Type 1: These respondents wish to influence the survey outcome correlated with their attitudes. They best respond to Type 0, assuming that 
\begin{enumerate}
\item All other respondents are of Type 0.
\item The estimator (designer) is only aware of Type 0 respondents.
\end{enumerate}
\item Type 2: These respondents have a higher level of strategic thinking and behave as the best response to a mix of Types 0 and 1, assuming that the designer (estimator) perceives the responses as coming from a distribution of these lower types. 
\end{itemize}

The designer of the survey is aware of the existence of these types of respondents as well as their true statistics. The question explored in this paper is: What is the designer's optimal ``de-biasing" procedure, i.e, optimally (in Bayesian sense) estimating the unbiased scores that reflect the true public reception of the plastic product?

We approach this problem via the recently introduced strategic quantization framework, see \cite{akyol2023isit}, which is a special case of the information design problem in Economics.\footnote{Throughout the paper, we use the terms quantizer and classifier interchangeably.}This class of problems, notable studied by \cite{rayo2010optimal,kamenica2011bayesian} explore the use of information by an agent (sender) to influence the action taken by another agent (receiver), where the aforementioned action determines the payoffs for both agents. Our prior work explored strategic quantization problem settings where the sender and the receiver were assumed to be fully rational agents. In this paper, we extend our strategic quantization work to settings with boundedly-rational sender (quantizer), via employing the cognitive hierarchy model of \cite{camerer2004cognitive}.

Throughout this paper, we focus on the quadratic distortion measures. Particularly,  the senders observe a two-dimensional source $(X, S)  \sim f_{X,S}(\cdot,\cdot)$ with a known joint density function over $X$ and $S$, where $X$ and $S$ can be interpreted as the state and bias variables. 
There are two types of strategic senders, both trying to minimize $\mathbb \{(X+S-\hat X)^2\}$, with different assumptions on the estimator (receiver). Type 1 strategic users assume the estimator is simply nonstrategic (which is the best response to Type 0). Type 2, in turn, assumes that the decoder is aware of a mix of Type 0 and Type 1 senders. 

The receiver's objective is to estimate the true state in the minimum mean squared error (MME) sense, i.e., the receiver minimizes $\eta(x,y)=(x-y)^2$ by choosing an action $\hat X$ which is the optimal MMSE estimate of $x$ given the quantization index from the sender $y= Q(x, s)$, hence $\hat X= \mathbb E\{X|Y=y\}$. In sharp contrast with the conventional quantization problem where the sender chooses $Q$ that minimizes $\mathbb E \{ (X-\hat X)^2\}$, in this setting the sender's choice of quantization mapping $Q$ minimizes a biased estimate, i.e.,  $\mathbb E \{ (X+S-\hat X)^2\}$. The objectives and the source distribution are common knowledge, available for all agents. 
We note that similar signaling problems with quadratic measures have been analyzed in the  Economics literature, see e.g., 
\cite{crawford1982strategic,fischer2000reporting,benabou2006incentives}.

This paper is organized as follows: In Section \RNum{2} we present the problem formulation. In Section \RNum{3},  we present a gradient-descent based algorithm to compute the classifier implemented by the boundedly rational agent. We provide numerical results
in Section \RNum{4}, and conclude in Section \RNum{5}. 

\begin{figure*}
    \centering
    \includegraphics[width=17cm]{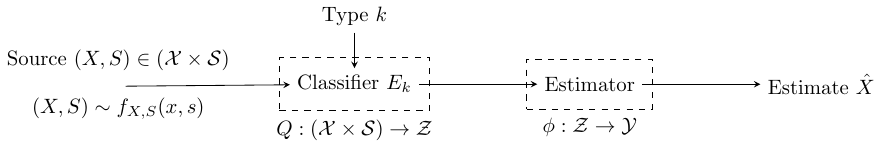}
    \caption{Communication diagram: with probability $p_k$, $E_k$ sends a message  
 $Z$ which is a function of the source $(X,S)$ over a noiseless channel}
    \label{fig:comm_theta_noiseless}
\end{figure*}
\section{Preliminaries}
\subsection{Notation}
In this paper, random variables are denoted using capital letters (say $X$), their sample values with respective lowercase letters ($x$), and their alphabet with respective calligraphic letters ($\mathcal{X}$). Vectors are denoted in bold font. The set of real numbers is denoted by $\mathbb{R}$. The alphabet, $\mathcal{X}$, can be finite, infinite, or a continuum, like an interval $[a,b]\subset \mathbb R$. The 
2-dimensional jointly Gaussian probability density function with mean $\begin{bmatrix}
    t_1 & t_2
\end{bmatrix}'$ and respective variances $\sigma_1^2,\sigma_2^2$ with a correlation $\rho$ is denoted by 
$\mathcal{N}\bigg (\begin{bmatrix}t_1\\ t_2    \end{bmatrix},\begin{bmatrix}
       \sigma_1^2  & \sigma_1 \sigma_2 \rho \\ \sigma_1 \sigma_2 \rho & \sigma_2^2 
    \end{bmatrix}\bigg ), 0\leq \rho <1$, $t_1,t_2\in \mathbb{R}$.  The expectation operator is written as $\mathbb{E}\{\cdot\}$. The operator $|\cdot |$ denotes the absolute value if the argument is a scalar real number and the cardinality if the argument is a set. 

\section{Problem Formulation}
Consider the following classification problem:  Three classifiers (senders), $E_k,k\in [0:2]$ each with a probability  $p_k$ of being chosen to send the message observe realizations of the two sources $X\in \mathcal{X}\subseteq  [a_X,b_X],S \in \mathcal{S}\subseteq [a_{S},b_{S}]$, $a_X,b_X,a_{S},b_{S} \in \mathbb{R}$ with joint probability density $(X,S)\sim f_{X,S}(\cdot,\cdot)$. The chosen classifier $E_k$ maps $(X,S)$ to a message $Z\in \mathcal{Z}$, where $\mathcal{Z}$ is a set of discrete messages with a cardinality constraint $|\mathcal{Z}|\leq M$  using a non-injective mapping $Q^k:(\mathcal{X}\times \mathcal{S})  \rightarrow \mathcal{Z}$.
After receiving the message $Z$, the receiver applies a mapping $\phi:\mathcal{Z}\rightarrow \mathcal{Y}$ on the message $Z$ and takes an action $Y=\phi(Z) $.

The set $\mathcal{X}$ is divided into mutually exclusive and exhaustive sets by each classifier $E_k$ as $\mathcal{V}_{1}^k,\mathcal{V}_{2}^k,\ldots,\mathcal{V}_{M}^k$. 
Let the marginal probability density function of $X$ be $f_X(x)$.

The probability $p_k$ of sender $E_k$ being chosen follows a normalized Poisson distribution
\begin{equation}
    p_k = \frac{e^{\lambda} \frac{\lambda^k}{k!}}{\sum_{i=0}^2e^{\lambda} \frac{\lambda^i}{i!}}.
\label{eqn:p_act}
\end{equation}

The distortion of the senders $E_1,E_2$ are $\eta_E^k(x,s,y)=(x+s-y)^2, k\in[1:2]$, that of sender $E_0$ is $\eta_E^0(x,s,y)=(x-y)^2$, and that of the receiver is $\eta_D(x,y)=(x-y)^2$.  
Let $y_m^{(k)}$ be the estimate that $E_k$ assumes are optimized by the receiver with respect to the respective perceived receiver distortion $D_D^{k}$, obtained by enforcing KKT conditions of optimality, $\partial D_D^{k}/\partial y_m^{(k)}=0$.  We consider three senders with hierarchical cognitive types and define the senders' and their respective perceived receiver distortions as $D_E^k,D_D^k,k\in[0:2]$ below:
\begin{enumerate}
    \item Non-strategic sender $E_0$: similar to level $L_0$ cognitive type, the sender assumes all senders are of type $E_0$, and that the decoder assumes all senders are of type $E_0$. Sender $E_0$ considers the receiver's distortion as the same as the sender's, $D_E^0=D_D^0$ (provides the information required by the receiver honestly)
    \begin{equation}
        D_D^0=\underset{m=1}{\overset{M}{\sum}} \mathbb{E}_S\{\eta_D(X,S,y_m^{(0)})|X\in \mathcal{V}_{m}^0\}.\nonumber
    \end{equation}
   \item Level-1 strategic sender $E_1$:  similar to level $L_1$ cognitive type, the sender assumes all other senders are of type $E_0$ and that it is uniquely of type $E_1$. The sender assumes the receiver thinks that all sender types are $E_0$, i.e., $D_D^1=D_D^0$, which results in the estimates perceived by $E_1$, $\mathbf{y}^{(1)}=\mathbf{y}^{(0)}$.
   \item Level-2 strategic sender $E_2$: similar to level $L_2$ cognitive type, the sender assumes the other senders are of lower cognitive levels and are Poisson distributed with a probability mass function $\mathbf{p}'=[p_0',p_1']$,
   \begin{equation}
       p_k'= \frac{e^{\lambda} \frac{\lambda^k}{k!}}{\sum_{i=0}^1e^{\lambda} \frac{\lambda^i}{i!}}, \nonumber
   \end{equation}
   for $E_k,k\in[0:1]$, respectively. Note that  
   this perceived probability mass function $\mathbf{p}'$ is not the actual statistics of the population, $\mathbf{p}'\neq \mathbf{p}$.
   The sender assumes the receiver is aware only of the types $E_0$ and $E_1$ and its perceived probability mass function $p_0'$ and $p_1'$. Sender $E_2$'s  perceived receiver distortion,
    \begin{align}
    D_D^2 &=\underset{i=0}{\overset{1}{\sum}}p_i'  \underset{m=1}{\overset{M}{\sum}} \mathbb{E}_{S}\{\eta_D(X,S,y_m^{(2)})|X\in \mathcal{V}_{m}^i\}.  
\nonumber
\end{align}
\end{enumerate}
The encoder distortions for each type $k\in[0:2]$,
\begin{equation}
     D_E^k= \underset{m=1}{\overset{M}{\sum}} \mathbb{E}_S\{\eta_E^k(X,S,y_m^{(k)})|X\in \mathcal{V}_{m}^k\}. \nonumber
 \end{equation}

 The receiver's distortion is given by
 \begin{equation}
     D_D^* =\underset{\mathbf{y}}{\min}\underset{i=0}{\overset{2}{\sum}} \underset{m=1}{\overset{M}{\sum}} p_i \mathbb{E}_{S}\{\eta_t^i(X,S,y_m|X\in \mathcal{V}_{m}^i\} ,\nonumber
 \end{equation}
  and $\mathbf{y}$ that minimizes the above expression is the actual receiver's action, $\mathbf{y}^*$.

Each sender type $E_k,k\in[1:2]$ optimizes their classifiers $Q^k$ with respect to their own distortion $D_E^k$, assuming the receiver is aware of only $E_i,i<k$ sender types. Sender $E_k,k\in[0:2]$ designs $Q^k$
\textit{ex-ante}, i.e., without the knowledge of the realization of $(X,S)$, using only the objectives $D_E^k$ and $D_D^k$, and the statistics of the source $f_{X,S}(\cdot,\cdot)$. 

The receiver is fully rational and has full information about the classification setup. The shared prior ($f_{X,S}$), the probability mass function over the sender types ($\mathbf{p}=[p_0,p_1,p_2]$) and the mappings ($\mathbf{Q}=\{Q^k,k\in[0:2]\}$) are known to the receiver.  The problem is to design the classifiers $\mathbf{Q}$ for the equilibrium, i.e., each sender type $E_k$ minimizes its own objective, assuming that the receiver minimizes its corresponding perceived objective $D_D^k$. This classification problem is given in Fig. \ref{fig:comm_theta_noiseless}. Since the senders choose the classifiers $\mathbf{Q}$ first, followed by the receiver choosing the perceived estimates ($\mathbf{y}^{(k)},k\in[0:2]$), we look for a Stackelberg equilibrium.  

The classifier design involves computing classifiers for each realization of $S$ by classifier $i$ as $\mathcal{U}_{s,m}^i,s\in \mathcal{S}$, where $\underset{s\in \mathcal{S}}{\cup}\mathcal{U}_{s,m}^i = \mathcal{V}_m^i $.
Throughout this paper, we make the following ``monotonicity'' assumption on the sets $\{\mathcal \mathcal{U}_{s,m}^i\}$.
\begin{assumption}
\label{as1}
$\mathcal \mathcal{U}_{s,m}^i$ is convex for all $m\in [1:M],s\in \mathcal{S},i\in [0:2]$. 
\end{assumption}
That is, $\mathcal{U}_{s,m}^i=[q_{s,m}^i,q_{s,m+1}^i],q_{s,m}^i<q_{s,m+1}^i$. Then, $Q^i=\{\mathbf{q}_{s}^i,s\in \mathcal{S}\}$, where $\mathbf{q}_s^i=[q_{s,0}^i,\ldots, q_{s,M+1}^i]$.

Since $E_0$'s distortion function $\eta_E^0(x,s,y)=(x-y)^2$ is not a function of $s$, $Q^0:(\mathcal{X} \times \mathcal{S}) \rightarrow \mathcal{Z}$ simplifies to $Q^0:\mathcal{X}\rightarrow \mathcal{Z}$. Let $\mathbf{q}_s^0=\mathbf{n}=[n_0,\ldots,n_M]$ for all $s\in \mathcal{S}$.    $E_0$ responds honestly, $D_D^0=D_E^0$ (equivalent to the non-strategic classification setting), hence its classifier $\mathbf{n}$ and perceived estimates $\mathbf{y}^{(0)}$ are,
\begin{equation}
    D_E^0= \underset{m=1}{\overset{M}{\sum}} \underset{n_{m-1}}{\overset{n_m}{\int}}(x-y_m^{(0)})^2 f_{X}(x) \mathrm d x,\nonumber\quad  y_m^{(0)} =\mathbb{E}\{X|x\in \mathcal{V}_{m}^0\}.
\end{equation}

 $E_1$ assumes that all other senders are of type $E_0$ and that the receiver views all senders as type $E_0$, i.e., $E_1$'s perceived estimates $y_m^{(1)}=y_m^{(0)}$, since $D_D^1=D_D^0$.

 $E_2$ assumes the other agents are of types $E_0$ and $E_1$ with probability mass function $\mathbf{p}'$, and that the receiver perceives the agents as the same, of types $E_0,E_1$ with a probability mass function $\mathbf{p}'$,
\begin{equation}
    D_D^2 = \underset{i=0}{\overset{1}{\sum}}p_i' \underset{m=1}{\overset{M}{\sum}}  \underset{a_{S}}{\overset{b_{S}}{\int}}\underset{q_{s,m-1}^i}{\overset{q_{s,m}^i}{\int}}(x-y_{m}^{(2)})^2 f_{X,S}(x,s) \mathrm d x \mathrm d s\nonumber,
\end{equation}
resulting in $E_2$'s perceived estimates $y_m^{(2)}$,
\begin{align}
    y_{m}^{(2)} = \frac{\underset{i=0}{\overset{1}{\sum}}p_i' \underset{a_{S}}{\overset{b_{S}}{\int}}\underset{q_{s,m-1}^i}{\overset{q_{s,m}^i}{\int}}x  f_{X,S}(x,s) \mathrm d x \mathrm d s}{\underset{i=0}{\overset{1}{\sum}}p_i'\underset{a_{S}}{\overset{b_{S}}{\int}}\underset{q_{s,m-1}^i}{\overset{q_{s,m}^i}{\int}}  f_{X,S}(x,s) \mathrm d x \mathrm d s}\label{eqn:rec_enc2}.
\end{align}

The classifier distortions for $E_k,k\in[1:2]$ are given by
\begin{equation}
       D_E^k = \underset{m=1}{\overset{M}{\sum}} \underset{a_{S}}{\overset{b_{S}}{\int}}\underset{q_{s,m-1}^k}{\overset{q_{s,m}^k}{\int}}(x+s-y_m^{(k)})^2 f_{X,S}(x,s) \mathrm d x \mathrm d s  \label{eqn:DEk}.
\end{equation}

The receiver's distortion and estimates are
\begin{align}
     D_D^* = \underset{i=0}{\overset{2}{\sum}}p_i \underset{m=1}{\overset{M}{\sum}}  \underset{a_{S}}{\overset{b_{S}}{\int}}\underset{q_{s,m-1}^i}{\overset{q_{s,m}^i}{\int}}(x-y_{m}^*)^2 f_{X,S}(x,s) \mathrm d x \mathrm d s\label{eqn:decoderdist_act} ,
\end{align}
\begin{align}
    y_m^* = \frac{\underset{i=0}{\overset{2}{\sum}}p_i \underset{a_{S}}{\overset{b_{S}}{\int}}\underset{q_{s,m-1}^i}{\overset{q_{s,m}^i}{\int}}x  f_{X,S}(x,s) \mathrm d x \mathrm d s}{\underset{i=0}{\overset{2}{\sum}}p_i \underset{a_{S}}{\overset{b_{S}}{\int}}\underset{q_{s,m-1}^i}{\overset{q_{s,m}^i}{\int}}  f_{X,S}(x,s) \mathrm d x \mathrm d s}\label{eqn:receiver_act} .
\end{align}

 The classifers implemented by $E_0,E_1,E_2$ are as follows:
 \begin{enumerate}
     \item $E_0$ implements a non-strategic (classical) classifier $\mathbf{n}$ for the given density of $X$, $f_X(\cdot)$.
     \item $E_1$ implements a nearest neighbor classifier for $X+s$ with respect to $\mathbf{y}^{(0)}$. Since $E_1$ assumes that the receiver views all senders as type $E_0$,
     the estimates perceived by $E_1$ is the non-strategic estimates $\mathbf{y}^{(1)}=\mathbf{y}^{(0)}$. Minimizing $D_E^1$ with $\mathbf{y}^{(0)}$ estimates results in a nearest neighbor classifier, which is the non-strategic classifier shifted by $s$ for each realization $s\in \mathcal{S}$, $\mathbf{q}_s^1=\mathbf{n}-s$.
     \item $E_2$ implements the classifier minimizing (\ref{eqn:DEk}) with the given source probability density function $f_{X,S}(x,s)$ assuming receiver estimates are given by (\ref{eqn:rec_enc2}).
 \end{enumerate}

\section{Design}
 In this section, we present our gradient-descent based algorithm for the optimization of $Q^2$.

In \cite{anand2024cdc}, we proposed a gradient-descent based algorithm to solve the problem of quantization of a 2-dimensional source $(X,S)$ by extending our algorithm in \cite{akyol2023isit} for a scalar source to the 2-dimensional setting by a simple method of computing quantizers for each value of $s\in \mathcal{S}$.

Here, we use similar methods as in \cite{anand2024cdc}, also using the known classifiers for $E_0,E_1$,  we perform gradient descent optimization with the objective as $D_E^2$ optimized over $Q^2$, assuming the estimates $\mathbf{y}^{(2)}$ are optimized for $D_D^2$.  Although the sender's objective $D_E^2$ depends on receiver estimates $\mathbf{y}^{(2)}$, since $\mathbf{y}^{(2)}$ is a function of $Q^2$, the optimization can be implemented as a function of solely $Q^2$.

\begin{figure}
    \centering
    \includegraphics[width=6cm]{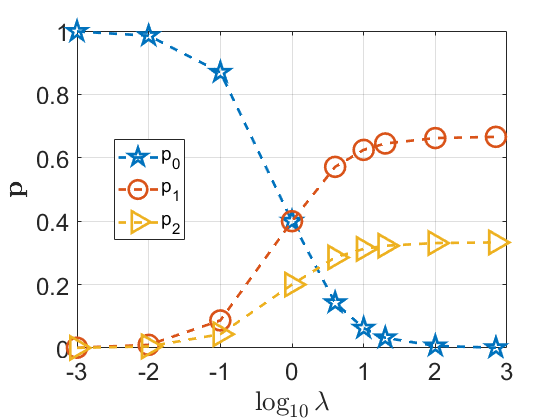}
    \caption{Probability mass function $\mathbf{p}$ with respect to $\lambda$.}
    \label{fig:population}
\end{figure}
\begin{figure*}[htb]
\begin{minipage}[b]{0.32\linewidth}
  \centering
  \centerline{\includegraphics[width=5.4cm]{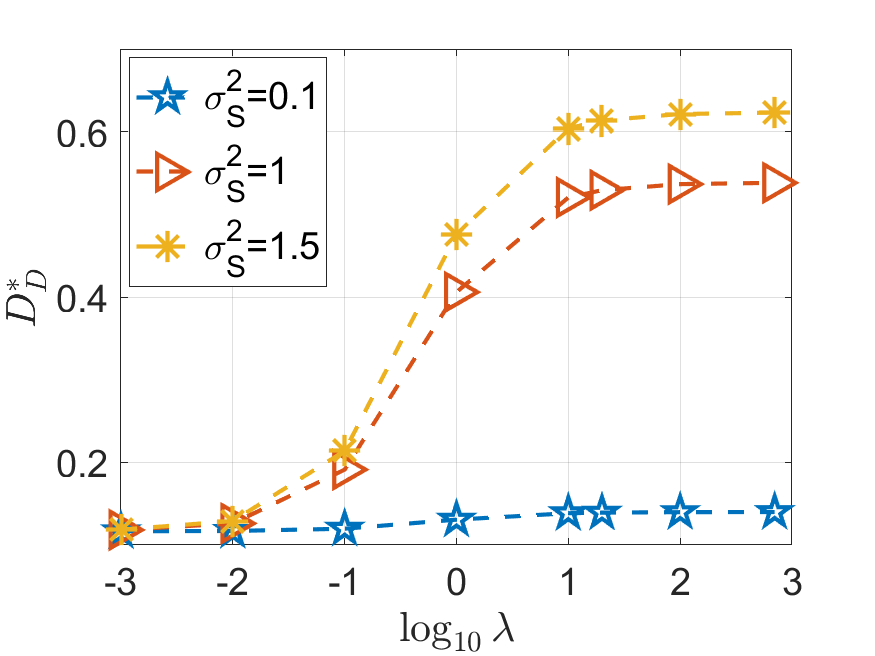}}
  \centerline{(a) $\rho=0.1$}
\end{minipage}
\hfill
\begin{minipage}[b]{.32\linewidth}
% \label{fig:}
  \centering
  \centerline{\includegraphics[width=5.4cm]{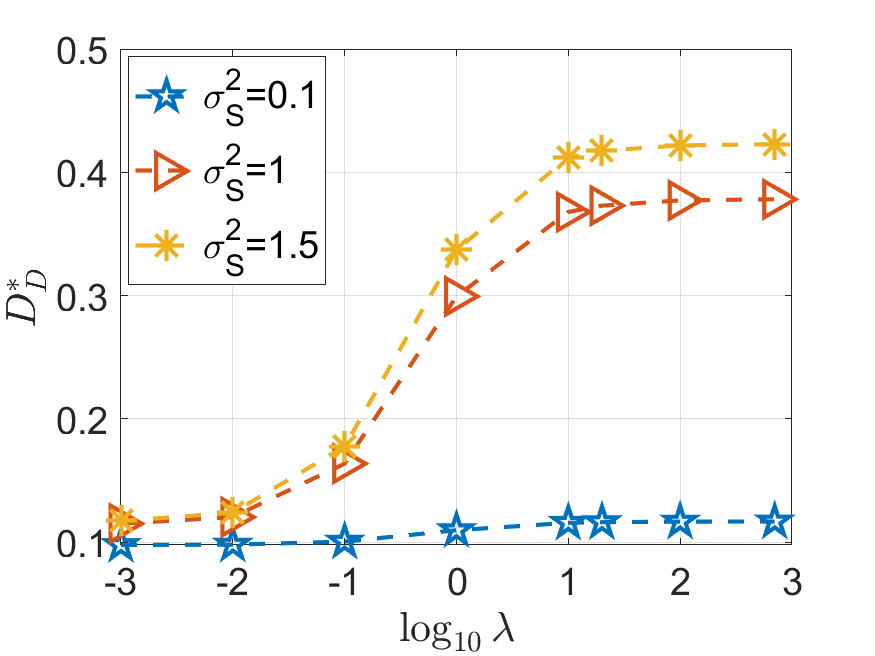}}
  \centerline{(b) $\rho=0.5$}
\end{minipage}
\hfill
\begin{minipage}[b]{.32\linewidth}
% \label{fig:}
  \centering
  \centerline{\includegraphics[width=5.4cm]{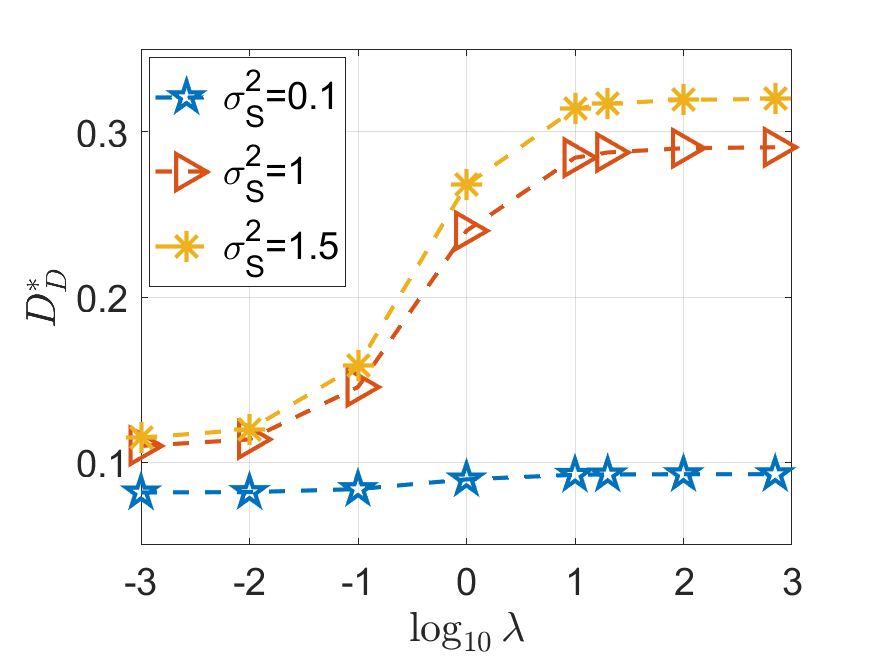}}
  \centerline{(c) $\rho=0.7$}
\end{minipage}
\caption{Receiver distortion $D_D^*$ for  $M=4$ classification of $(X,S) \sim \mathcal{N}\bigg(\begin{bmatrix}
    0 \\ 0
\end{bmatrix}, \begin{bmatrix}
    1 & \sigma_S \rho  \\  \sigma_S \rho & \sigma_S^2
\end{bmatrix} \bigg)$ for a given correlation $\rho$ value with respect to bias variance $\sigma_S^2$.}
\label{fig:wrtrho}
\end{figure*}
\begin{algorithm}[h!tbp]
\caption{Proposed strategic quantizer design} 
 Parameters: $\epsilon,\lambda$\\
 Input: $f_{X,S}(\cdot,\cdot),\mathcal{X},\mathcal{S},M,\eta_E^2,\eta_D^2,Q^0,Q^1,\mathbf{p},\mathbf{p}'$\\
 Output: $Q^{2}$, $\mathbf{y}^{(2)}$, $\mathbf{y}^{*}$, $D_E^2$, $D_D^*$\\
 Initialization: assign a set of monotone $\{\mathbf{q}_{s}^2\}_0$ randomly, compute associated sender distortion $D_E^2(0)$, set iteration index $j=1$;\\
\While{ $\Delta D >\epsilon $ or until a set amount of iterations} {
     compute the gradients $\{\partial D_E^2 /\partial \mathbf{q}_{s}^2\}_j$,\;\\
    compute the updated classifier  $\{\mathbf{q}_{s}^2\}_{j+1} \triangleq \{\mathbf{q}_{s}^2 \}_{j}- \lambda  \{\partial D_E^2 /\partial \mathbf{q}_{s}^2\}_j$ for $s \in \mathcal{S}$, \\
    compute estimates  $\mathbf{y}(\{\mathbf{q}_{s}^2\}_{j+1})$ via (\ref{eqn:rec_enc2}),\\
    compute sender distortion $D_E^2 (j+1)$ associated with classifier  $\{\mathbf{q}_{s}^2\}_{j+1}$ and estimates $\mathbf{y}(\{\mathbf{q}_{s}^2\}_{j+1})$ via (\ref{eqn:DEk}),\\
    compute  $\Delta D =D_E^2 (j)-D_E^2 (j+1)$. 
   }  
\Return classifier $Q^{2}=\{\mathbf{q}_{s}^2\}_{j+1}$, $E_2$'s perceived receiver estimates $\mathbf{y}^{(2)}=\mathbf{y}(Q^{2})$, actual receiver estimates $\mathbf{y}^*$, sender distortion $D_E^2$ computed for the bounded rational optimal classifiers $\mathbf{Q}=\{Q^k,\mathbf{y}^{(k)},k\in[0,2]\}$ with the perceived receiver estimates $\mathbf{y}^{(2)}$, and receiver distortion $D_D^*$ computed for the bounded rational optimal classifiers $\mathbf{Q}$ with the actual receiver estimate $\mathbf{y}^{*}$ via  (\ref{eqn:receiver_act}), (\ref{eqn:DEk}), (\ref{eqn:decoderdist_act}).\; 
\end{algorithm}

Like any gradient-descent-based algorithm, the proposed method may get stuck at a poor local optimum, which we resolve with a simple remedy by performing gradient descent with multiple initializations and choosing the best local optimum among them. 
A sketch of the proposed method is summarized in Algorithm 1. The MATLAB codes are provided at 
\url{https://github.com/strategic-quantization/bounded-rationality} for research purposes.

\section{Numerical Results}

\label{sec:results}
We consider a jointly Gaussian 2-dimensional source $$(X,S) \sim \mathcal{N}\bigg(\begin{bmatrix}
    0 \\ 0
\end{bmatrix}, \begin{bmatrix}
    1 & \sigma_S \rho \\ \sigma_S \rho & \sigma_S^2
\end{bmatrix} \bigg)$$ and present results  for different settings with parameters bias variance  $\sigma_S^2\in \{0.1,1,1.5\}$, correlation $\rho \in \{ 0.1, 0.5,0.7\}$, and  $\mathbf{p}$ following (\ref{eqn:p_act}) with  $\lambda\in [0.001,700]$ for an $M=4$ classifier. The probability mass function over the sender types $\mathbf{p}$ for different values of $\lambda$ is plotted in Fig. \ref{fig:population}. 
We consider only positive correlation since people's preferences are positively correlated with their opinions; for instance, both climate activists and climate change deniers try to bias their classification towards the extremes on their side. We plot the receiver's distortion $D_D^*$ in Fig.
\ref{fig:wrtrho} for given values of $\sigma_S^2$ and $\rho$, respectively.

We now interpret our results in terms of the impact of different parameters on the receiver's estimation.

\begin{figure}
    \centering
    \includegraphics[width=6cm]{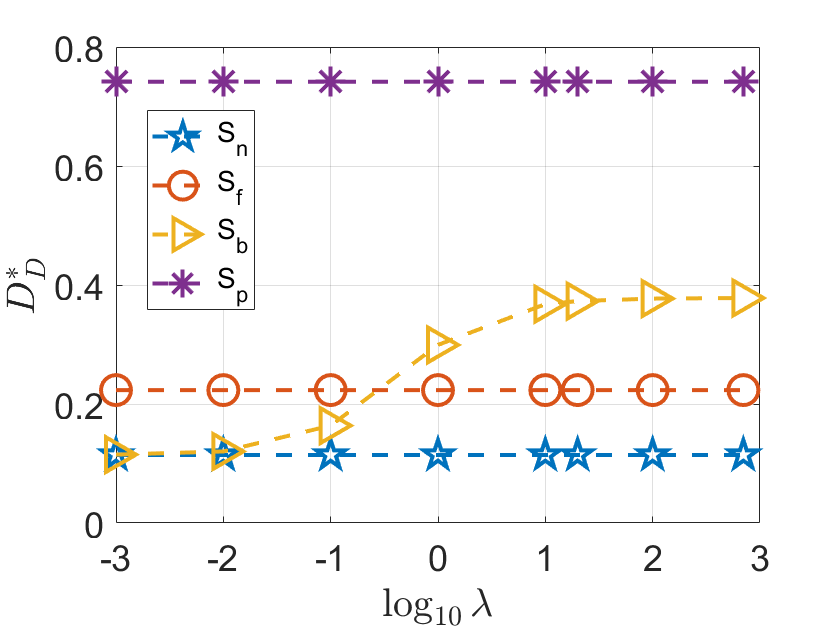}
    \caption{Receiver distortion $D_D^*$ for $M=4$ classification of  $(X,S) \sim \mathcal{N}\bigg(\begin{bmatrix}
    0 \\ 0
\end{bmatrix}, \begin{bmatrix}
    1 & 0.5 \\ 0.5 & 1
\end{bmatrix} \bigg)$  for a full information fully rational estimator with four different types of senders.}
    \label{fig:comparison}
\end{figure}
We observe from Fig. \ref{fig:wrtrho} that the correlation between the state and bias variables $\rho$ does not change the receiver distortion trends.
\subsection{Impact of cognitive parameter $\lambda$}
From Fig. \ref{fig:population} we note that as $\lambda\rightarrow 0$, the population mostly consists of level-0 cognitive level. As $\lambda$ increases, the population shifts towards higher cognitive types, and we expect the receiver distortion to increase with $\lambda$, as we observe in Fig.
\ref{fig:wrtrho}.
For $\lambda \rightarrow 0$, the receiver distortion does not change significantly with varying bias variance $\sigma_S^2$ since the population is mostly of level-0 type, and they respond honestly. 
For $\lambda>100$, the statistics of the population remain fairly constant, and hence the receiver distortion varies negligibly.

\subsection{Impact of varying $\sigma_S^2$}
For a given correlation $\rho$, we observe in Fig. \ref{fig:wrtrho} that as $\sigma_S^2$ decreases, the receiver distortion decreases. As the variance of the bias $\sigma_S^2$ decreases, the sender's opinions are closer to their true value, and the objectives of the sender and the receiver become more aligned. 

When the bias is negligible ($\sigma_S\rightarrow 0$), the objectives of all the senders are similar, resulting in a negligible change in the receiver distortion with $\lambda$, which we observe in Fig. \ref{fig:wrtrho} for  $\sigma_S^2=0.1$.

\subsection{Comparison with different types of senders}
In Fig. \ref{fig:comparison}, the receiver distortion for the following four different types of senders is plotted for a specific setting with $\sigma_X^2=\sigma_{S}^2=1,\rho=0.5$:
\begin{enumerate}
    \item non-strategic ($S_n$): All agents are non-strategic and send their honest reply ($E_0$).
    \item full information ($S_f$): All agents are fully rational and have full information. The classifier here is that in \cite{anand2024cdc}. 
    \item bounded rational ($S_b$): The agents follow the setting described in this paper. 
    \item partially-strategic ($S_p$): All agents minimize $\mathbb{E}\{(X+S-Y)^2\}$, but they assume the receiver is not strategic and hence implements a naive estimator, $\mathbf{y}^{(0)}$. The classifier is the same as that for $E_1$.
\end{enumerate}
The receiver is fully rational with full information about the type of sender, the source distribution, and sender and receiver objectives. 

As expected, the non-strategic sender results in the lowest receiver distortion. For negligible $\lambda$, the population is mostly of level-0 cognitive type, as mentioned before. Since they respond honestly, $S_b$ is closer to the non-strategic value as $\lambda \rightarrow 0$. 

Although we expect that $S_f$ results in maximizing the receiver distortion among the above four senders, we observe from Fig. \ref{fig:comparison} that the receiver may prefer a fully rational sender with full information to other types of strategic senders for $\lambda>\lambda_0$ for some $\lambda_0$.
We observe that the receiver benefits from a boundedly or fully rational setup compared to the setting where all senders are partially-strategic.

\section{Conclusions}
In this paper, we analyzed the problem of strategic classification of a 2-dimensional source $(X,S)$ with three types of senders with hierarchical cognitive types: level-0 non-strategic, level-1 strategic, and level-2 strategic, where each level assumes they are unique and that the other agents have lower cognitive levels.  We considered quadratic objectives for the sender and the receiver and extended our prior work on design, a gradient-descent based algorithm for a 2-dimensional source with a single type of fully rational sender with full information, to the bounded-rational setting considered in this paper.  The numerical results obtained via the proposed algorithm suggest several intriguing research problems that we leave as a part of our future work.

\label{sec:reference_examples}
\bibliographystyle{IEEEtran}
\bibliography{Ref1}
\end{document}